\documentclass[runningheads,a4paper]{llncs}

\usepackage{amssymb}
\usepackage{graphicx}
\usepackage[latin1]{inputenc}

\usepackage{url}
\urldef{\mailsa}\path|{alfred.hofmann, ursula.barth, ingrid.haas, frank.holzwarth,|
\urldef{\mailsb}\path|anna.kramer, leonie.kunz, christine.reiss, nicole.sator,|
\urldef{\mailsc}\path|erika.siebert-cole, peter.strasser, lncs}@springer.com|    
\newcommand{\keywords}[1]{\par\addvspace\baselineskip
\noindent\keywordname\enspace\ignorespaces#1}

\begin{document}

\mainmatter  

\title{Helping Domain Experts Build Phrasal Speech Translation Systems}

\titlerunning{Helping Domain Experts Build Speech Translation Systems}

%
%
\author{Manny Rayner 
\and Alejandro Armando \and Pierrette Bouillon\and Sarah Ebling \and Johanna Gerlach \and Sonia Halimi \and Irene Strasly\and Nikos Tsourakis\thanks{Medical translation work was supported by Geneva University's Innogap program. Work on sign language translation was supported by the Crédit Suisse, Raiffeisen, TeamCO and Max Bircher Foundations. We thank Nuance Inc. and the University of East Anglia for generously allowing us to use their software for research purposes.}}
\authorrunning{Helping Domain Experts Build Phrasal Speech Translation Systems}

\institute{University of Geneva, FTI/TIM, Switzerland\\
\url{http://www.unige.ch/traduction-interpretation/}\\
University of Zurich, Institute of Computational Linguistics, Switzerland\\
\url{http://www.cl.uzh.ch/}}

%
%

\toctitle{Title1}
\tocauthor{Title2}
\maketitle

\begin{abstract}
We present a new platform, ``Regulus Lite'', which supports rapid
development and web deployment of several types of phrasal speech
translation systems using a minimal formalism. A distinguishing
feature is that most development work can be performed directly by
domain experts. We motivate the need for platforms of this type and
discuss three specific cases: medical speech translation,
speech-to-sign-language translation and voice questionnaires. We
briefly describe initial experiences in developing practical systems.

\end{abstract}

\section{Introduction and motivation}

In this paper, we claim that there is a place for limited-domain
rule-based speech translation systems which are more expressive than
fixed-phrase but less expressive than general syntax-based transfer or
interlingua architectures. We want it to be possible to construct
these systems using a formalism that permits a domain expert to do
most of the work and immediately deploy the result over the web. To
this end, we describe a new platform, ``Regulus Lite'', which can be
used to develop several different types of spoken language translation
application.

A question immediately arises: are such platforms still relevant,
given the existence of Google Translate (GT) and similar engines? We
argue the answer is yes, with the clearest evidence perhaps coming
from medical speech translation. Recent studies show, unsurprisingly,
that GT is starting to be used in hospitals, for the obvious reason
that it is vastly cheaper than paid human interpreters
\cite{chang2014}; on the other hand, experience shows
that GT, which has not been trained for this domain, is seriously
unreliable on medical language.  A recent paper \cite{patil2014}
describes the result of a semi-formal evaluation, in which it was used to translate
ten text sentences that a doctor might plausibly say to a patient into 26 target
languages.  The bottom-line conclusion was that
the results were incorrect more than one time in three.

Doctors are thus with good reason suspicious about the use of
broad-coverage speech translation systems in medical contexts, and
the existence of systems like
MediBabble\footnote{\url{http://medibabble.com/}} gives further
grounds to believe that there is a real problem to solve
here. MediBabble builds on extremely unsophisticated translation
technology (fixed-phrase, no speech input), but has achieved
considerable popularity with medical practitioners.  In
safety-critical domains like medicine, there certainly seem to be many
users who prefer a reliable, unsophisticated system to an unreliable,
sophisticated one. MediBabble is a highly regarded app because the
content is well-chosen and the translations are known to be good, and
the rest is viewed as less important. The app has been constructed
by doctors; a language technologist's reaction is that even if GT may
be too unreliable for use in hospitals, one can hope that it is not
necessary to go back to architectures quite as basic as this. A
reasonable ambition is to search for a compromise which retains the
desirable property of producing only reliable output prechecked by
professional translators, but at the same time supports at least some
kind of productive use of language, and also speech recognition.

A second type of application which has helped motivate the development
of our architecture is speech-to-sign-language translation. Sign languages
are low-resource, a problem they share with many of the target languages
interesting in the context of medical speech translation.
In addition, since they are non-linear, inherently relying on multiple parallel
channels of communication including hand movement, eye gaze, head tilt
and eyebrow inflection \cite{Neidle2000}, it is not possible to
formalise translation as the problem of converting a source-language
string into a target-language string. It is in principle feasible to
extend the SMT paradigm to cover this type of scenario, but currently
available mainstream SMT engines do not do so.  As a result, most
previous SMT approaches to sign language machine translation, such as
\cite{stein-schmidt-ney-2012} and \cite{morrissey-2008}, have used
unilinear representations of the sign languages involved.  
If we want to build sign-language translators
which can produce high-quality output in the short-term, rule-based
systems are a logical choice.

A third application area where this kind of approach seems 
appropriate is interactive multilingual questionnaires. Particularly
in crisis areas, it is often useful for personnel in the field to be
able to carry out quick surveys where information is elicited from
subjects who have no language in common with the interviewer
\cite{Salihu2013}.
Again,
applications of this kind only need simple and rigid coverage, but
accurate translation and rapid deployability are essential,
and practically interesting target languages are often underresourced.

In the rest of the paper, we describe Regulus Lite, showing how it can
be used as an economical tool for building spoken language translation
applications at least for the three domains we have just
mentioned. The main focus is application content
development. \S\ref{Section:platform} gives an overview of the
platform and the rule formalism. \S\ref{Section:applications} presents
specific details on medical speech translation, sign language
translation and voice questionnaires, and briefly sketches the initial
applications. \S\ref{Section:PrototypesEvaluations} presents some
initial evaluation results for the voice questionnaire app, currently
the most advanced one. The final section concludes.

\section{The platform}
\label{Section:platform}



The Regulus Lite platform supports rapid development and web
deployment for three types of small to medium vocabulary speech
translation applications: plain translation, sign language translation,
and voice questionnaires. We briefly describe each of these:

\begin{description}

\item[Plain translation] The simplest case: the source language user speaks and the system
displays its understanding of what the user said (a paraphrase of what was recognised).
If the source language user approves the paraphrase, a target language translation is
produced. 

\item[Sign language translation] Similar to plain translation, but the output is rendered
in some form of sign language, using a signing avatar.

\item[Voice questionnaires] The content is organized as a form-filling questionnaire,
where the interviewer poses the questions in spoken form,
after which they are translated into the target language and presented to the subject. 
There are typically many possible questions for each field in the questionnaire.
The subject responds by pressing one of a question-dependent set of buttons,
each of which is labelled with a possible answer.

\end{description}


A basic assumption is that the content will be in the form of flat
phrasal regular expression grammars. Reflecting this, content is
specified using two basic constructions, {\tt TrPhrase} (phrases) and
{\tt TrLex} (lexical items). Each construction combines one or more
{\tt Source} language patterns and at most one {\tt Target} language
result for each relevant target language, and indicates that the {\tt Source}
line can be translated as the {\tt Target}. A trivial example\footnote{The notation has been changed
  slightly for expositional purposes.} might be
\begin{verbatim}
TrPhrase $$top
Source ( hello | hi )
Target/french Bonjour
EndTrPhrase
\end{verbatim}
A slightly more complex example, which includes a {\tt TrLex}, might be
\begin{verbatim}
TrPhrase $$top
Source i ( want | would like ) $$food-or-drink ?please
Source ( could | can ) i have  $$food-or-drink ?please
Target/french je voudrais $$food-or-drink s'il vous plaît
EndTrPhrase

TrLex $$food-or-drink source="a (coca-cola | coke)" french="un coca" 
\end{verbatim}
Here, the variable {\tt \$\$food-or-drink} in the first rule indicates a phrase that is
to be translated using the second rule.

In order to decouple the source language and target language
development tasks, {\tt TrPhrase} and {\tt TrLex} units are split into
pieces placed in separate language-specific files, one for the source
language and one for each target language. The connecting link is
provided by a canonical version of the source language text (the
portions marked as {\tt Target/english} or {\tt english=}). Thus the
{\tt TrPhrase} and {\tt TrLex} units above will be reconstituted from
the source-language (English) pieces
\begin{verbatim}
TrPhrase $$top
Source i ( want | would like ) $$food-or-drink ?please
Source ( could | can ) i have  $$food-or-drink ?please
Target/english i want $$food-or-drink please
EndTrPhrase

TrLex $$food-or-drink source="a (coca-cola | coke)" english="a coke"
\end{verbatim}
and the target language (French) pieces
\begin{verbatim}
TrPhrase $$top
Target/english i want $$food-or-drink please
Target/french je voudrais $$food-or-drink s'il vous plaît
EndTrPhrase

TrLex $$food-or-drink english="a coke" french="un coca"
\end{verbatim}
The development process starts with the source language developer
writing their piece of each unit, defining the application's
coverage.  A script then generates ``blank'' versions of the
target language files, in which the canonical source lines are filled
in and the target language lines are left empty; so the French
target language developer will receive a file containing
items like the following, where their task is to replace the question
marks by translating the canonical English sentences.
\begin{verbatim}
TrPhrase $$top
Target/english i want $$food-or-drink please
Target/french ?
EndTrPhrase

TrLex $$food-or-drink source="a coke" french="?"
\end{verbatim}
As the source language developer adds more coverage, the ``blank''
target language files are periodically updated to include relevant new
items.

The content can at any time be compiled into various pieces of runtime
software, of which the most important are an application-specific
grammar-based speech recogniser and a translation grammar;
the underlying speech recognition engine used in the implemented
version of the platform is Nuance Recognizer version 10.2.  These
generated software modules can be immediately uploaded to a webserver,
so that the system is redeployable on a time scale of a few minutes.
Applications can be hosted on mobile
platforms --- smartphones, tablets or laptops --- linked over a 3G
connection to a remote server, with recognition performed on the 
server \cite{FuchsEALREC2012}.
The deployment-level architecture of the platform is adapted from that
of the related platform described in \cite{RaynerEASlate2015Platform},
and offers essentially the same functionality.

\section{Types of application}
\label{Section:applications}

\subsection{Medical translation}

As already mentioned, medical speech translation is one of the areas
which most strongly motivates our architecture. Several studies,
including earlier projects of our own \cite{patil2014,Tsourakis2013}, suggest that
doctors are dubious about the unpredictability of broad-coverage SMT
systems and place high value on translations which have been
previously validated by professional translators. Other relevant
factors are that medical diagnosis dialogues are sterotypical and
highly structured, and that the languages which pose practical
difficulties are ones badly served by mainstream translation
systems. 

The practical problems arise from the fact that the Lite formalism
only supports regular expression translation grammars. 
The question
is thus what constituents we can find which it is safe always to
translate compositionally. It is clear that many constituents cannot
be treated in this way. Nonetheless, it turns out that enough
of them can be translated compositionally that the grammar
description is vastly more efficient than a completely enumerative
framework; most adjuncts, in particular PPs and
subordinate clauses, can be regarded as compositional,
and it is often possible to treat nouns and adjectives compositionally
in specific contexts. 

We are currently developing a prototype medical speech translator in a
collaboration with a group at Geneva's largest
hospital\footnote{Hôpitaux Universitaires de Genève}. Initial coverage
is organised around medical examinations involving abdominal pain,
with the rules loosely based on those developed under an earlier
project \cite{BouillonEA2008AMTA}.  Translation is from French to
Spanish, Italian and Arabic\footnote{Tigrinya will be added soon.}.  A typical source language rule
(slightly simplified for presentational purposes) is
\begin{verbatim}
TrPhrase $$top
Source ?$$PP_time la douleur est-elle ?$$adv $$qual ?$$PP_time
Source ?$$PP_time avez-vous ?$$adv une douleur $$qual
Source ?$$PP_time ?(est-ce que) la douleur est ?$$adv $$qual ?$$PP_time
Target/french la douleur est-elle ?$$adv $$qual ?$$PP_time
EndTrPhrase
\end{verbatim}
Here, the French {\tt Source} lines give different variants of {\em la
  douleur est-elle \$\$qual} (``Is the pain \$\$qual?''), for various
substitutions of the transfer variable {\tt \$\$qual} ({\em vive},
``sharp''; {\em difficile à situer}, ``hard to localize''; {\em dans
  l'angle costal}, ``in the intercostal angle'', etc). Each variant
can optionally be modified by an adverb ({\tt \$\$adv}) and/or a
temporal PP ({\tt \$\$PP\_time}).  Thus the questions covered will be
things like {\em avez-vous souvent une douleur vive le matin?} (``do
you often experience a sharp pain in the morning?'') As the rule
illustrates, there are typically many possible ways of formulating the
question, all of which map onto a single canonical version. The
target language translators work directly from the canonical version.

The current prototype represents the result of about one person-month
of effort, nearly all of which was spent on developing the source side
rules. Coverage consists of about 250 canonical patterns, expanding to
about 3M possible source side sentences; the source language
vocabulary is about 650 words. Creating a set of target language rules
only involves translating the canonical patterns, and is very quick;
for example, the rules for Italian, which were added at a late stage,
took a few hours. 

Speech recognition is anecdotally quite good: sentences which are
within coverage are usually recognised, and correctly recognised
utterances are always translated correctly.  The informal opinion of
the medical staff who have taken part in the experiments is that the
system is already close to the point where it would be useful in real
hospital situations, and clearly outperforms Google Translate within
its intended area of application. We are in process of organising a
first formal evaluation and expect to be able to report results in
2016.


\subsection{Sign language translation}



The rapidly emerging field of automatic sign language translation
poses multiple challenges
\cite{BraffortEA2005,CoxEA2002,EblingGlauert2013,elliott2000,Huenerfauth2006,Kennaway2002,MarshallEA2002,MazzeiEA2013,MorriseyEA2007,OngEA2005,SuWu2009}. An
immediate problem is that sign languages are very resource-poor. Even
for the largest and best-understood sign languages, ASL and Auslan,
the difficulty and expense of signed language annotation means there
is an acute shortage of available corpus data\footnote{The largest
  parallel corpus used in sign language translation that we know of
  has about 8\,700 utterances \cite{ForsterEA2014}.}; for most of the
world's estimated 120 sign languages \cite{zeshan-2012}, there are no
corpora at all. In addition, there are often no reliable lexica or
grammars and no native speakers of the language with training in
linguistics.

Sign languages also pose unique
challenges not shared with spoken languages. As already mentioned,
they are inherently non-linear; even though the most important
component of meaning is conveyed by the hands/arms (the
\emph{manual activity}), movements of the shoulders, head, and face (the \emph{non-manual components}) are also extremely important and are capable of assuming  functions at all linguistic levels \cite{crasborn-2006}. Commonly cited examples include the use of head shakes/eyebrow
movements to indicate negation and eye gaze/head tilt to convey
topicalization \cite{Neidle2000,JohnstonSchembri2007}.
Anecdotally, signers can to
some extent understand signed language which only uses hand movements,
but it is regarded as unnatural and can easily
degenerate into incomprehensibility \cite{TomaszewskiFarris2010};
quantitatively, controlled studies show that the absence of non-manual information in synthesized signing (sign language animation) leads to lower comprehension scores and lower subjective ratings of the animations \cite{kacorri-lu-huenerfauth-2013}.
In summary, it is unsatisfactory to model sign language translation with the
approximation most often used in practice: represent a signed
utterance as a sequence of ``glosses'' (identifiers corresponding to
hand signs), and consider the translation problem as that of finding a
sequence of glosses corresponding to the source language utterance
\cite{ebling-huenerfauth-2015}. This approximation is unfortunately necessary if
mainstream SMT engines are to be used.


For the above reasons and others, it is natural to argue that
current technology requires high-quality automatic sign language
translation to use rule-based methods in which signed utterances are
represented in nonlinear form \cite{Huenerfauth2006}. Our
treatment conforms to these intuitions and adapts them to the
minimalistic Lite framework. Following standard practice in the sign language linguistics
literature, a signed utterance is represented at the
linguistic level as a set of aligned lists, one for each parallel
output stream: at the moment, we use six lists respectively called
{\tt gloss} (hand signs), {\tt head} (head movements like nodding or
shaking), {\tt gaze} (direction of eye gaze), {\tt eyebrows} (raising
or furrowing of eyebrows), {\tt aperture} (widening or narrowing of
eyes) and {\tt mouthing} (forming of sound-like shapes with the
mouth). 

\begin{figure*}
\begin{center}
\begin{verbatim}
gloss      TRAIN1NORMAL   CE           GENEVE   ALLER     PAS      
head       Down           Down         Neutral  Neutral   Shaking  
gaze       Neutral        Down         Neutral  Neutral   Neutral  
eyebrows   FurrowBoth     FurrowBoth   Up       Up        Neutral  
aperture   Small          Small        Neutral  Wide      Neutral  
mouthing   Tr@            SS           Genève   Vais      Pas      
\end{verbatim}
\end{center}
\caption{Sign table representation of an utterance in Swiss French Sign Language
meaning ``This train does not go through Geneva''.}
\label{Figure:SignedExample}
\end{figure*}

The examples we show below are taken from our initial application,
which translates train service announcements from spoken French to
Swiss French sign language. A typical sign table is shown in
Figure~\ref{Figure:SignedExample};
translation from speech to sign is performed in three stages, with
sign tables like these acting as an intermediate representation or pivot. As before, the
first stage is to use speech recognition to produce a source language text 
string\footnote{This is a slight oversimplification; in actual fact,
recognition passes an n-best hypothesis list. The complications this introduces are
irrelevant in the present context.}. In the second, the source language string
is translated into a sign table. Finally, the sign table is translated into 
a representation in SiGML \cite{elliott2000}, which can be fed into a signing avatar; in 
the current version of the system, we use JASigning \cite{elliott2008}.
The image below shows the user interface. On the left, we
have, from top to bottom, the recognition result and the sign table;
on the right, the avatar, the avatar controls and the SiGML.

\vspace{5pt}
\includegraphics[width=11.5cm]{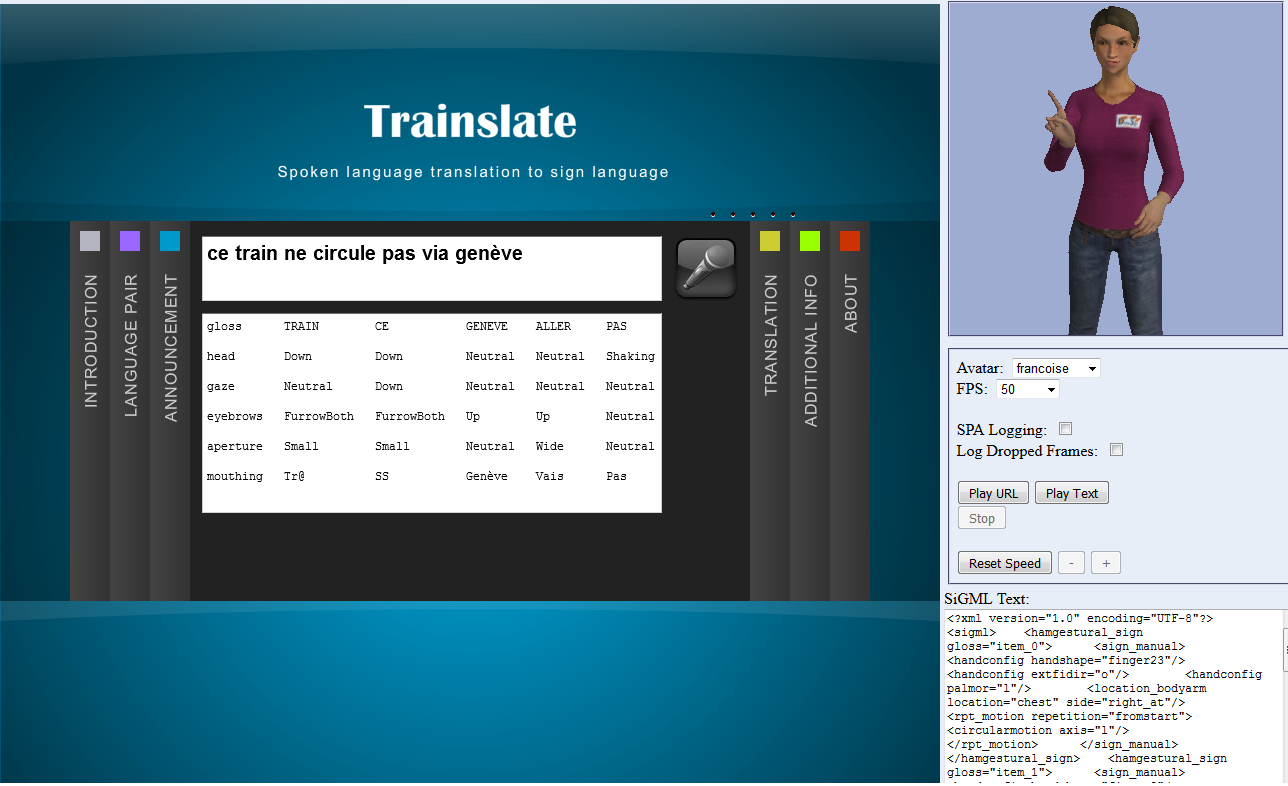}
\vspace{5pt}

The issues that are of interest here are concerned with the text to
sign table translation stage; once again, the central challenge is to
create a formalism which can be used by linguists who are familiar
with the conventions of sign language linguistics, but not necessarily
with computational concepts.  The formalism used is a natural
generalization of the one employed for normal text-to-text
translation; the difference is that the output is not one list of
tokens, but six aligned lists, one for each of the output streams.
For practical reasons, since correct alignment of the streams is
crucial, it is convenient to write rules in spreadsheets and use the
spreadsheet columns to enforce the alignment.

The non-obvious aspects arise from the fact that phrasal sign
translation rules in general fail to specify values for all the output
streams, with the values of the other streams being filled in by
phrases higher up in the parse tree. Figure~\ref{Figure:SignRules}
illustrates, continuing the example from the previous figure. The
lexical entry for {\em genève} only specifies values for
the {\tt gloss} and {\tt mouthing} lines.  When the rules are combined
to form the output shown in Figure~\ref{Figure:SignedExample}, the
value of {\tt eyebrows} associated with the sign glossed {\tt GENEVE}
is inherited from the phrase above, and thus becomes {\tt FurrowBoth}.

\begin{figure*}
\begin{center}
\begin{verbatim}
TrPhrase $$top  
Source ce train ne circule pas via $$station           
Target/gloss    TRAIN      CE         $$station    ALLER   PAS           
Target/head     Down       Down       Neutral      Neutral Shaking           
Target/gaze     Neutral    Down       Neutral      Neutral Neutral           
Target/eyebrows FurrowBoth FurrowBoth Up           Up      Neutral           
Target/aperture Small      Small      Neutral      Wide    Neutral           
Target/mouthing Tr@        SS         $$station    Vais    Pas           
EndTrPhrase                

TrLex $$station source="genève" gloss="GENEVE" mouthing="Genève"       
\end{verbatim}
\end{center}
\caption{Examples of top-level translation rule and lexical entry for the
train announcement domain. 
The rule defines a phrase of the form
{\em ce train ne circule pas via $\langle$station$\rangle$} (``this train does
not travel via $\langle$station$\rangle$''. The lexical entry
defines the translation for the name {\em genève} (``Geneva'').
Only gloss and mouthing forms are defined for the lexical item.}
\label{Figure:SignRules}
\end{figure*}

The process by which sign tables are translated into SiGML is
tangential to the main focus of this paper, so we content ourselves
with a brief summary.  The information required to perform the
translation is supplied by three lexicon spreadsheets, maintained by
the sign language expert, which associate glosses and other
identifiers appearing in the sign table with SiGML tags and strings
written in HamNoSys \cite{Prillwitz1989}, a popular notation for
describing signs. The rule compiler checks the spreadsheets for
missing entries, and if necessary adds new ``blank'' rows, using a
model similar to that described in \S\ref{Section:platform}.

\subsection{Voice questionnaires}



We have already touched on the special problems of interactive voice
questionnaires in the introduction. The overall intention is to 
add speech input and output capablities to the RAMP data gathering questionnaire
framework \cite{Salihu2013}. The questionnaire definition encodes a
branching sequence of questions, where the identity of the following
question is determined by the answer to the preceding one. The display
shows the person administering the questionnaire the field currently
being filled; they formulate a question and speak it in their own
language. In general, there are many questions which can be used to
fill a given field, and the interviewer will choose an appropriate one
depending on the situation. A basic choice, which affects most fields,
is between a WH and a Y/N question.  For example, if the interviewer
can see recently used cooking utensils in front of him, it is odd to
ask the open-ended WH-question ``Where do you do the cooking?''; a
more natural choice is to point and ask the Y/N confirmation question
``Is cooking done in the house?'' 

As usual, the app performs speech recognition, gives the interviewer
confirmation feedback, and speaks the target language translation if
they approve. It then displays a question-dependent set of answer
icons on the touch-screen. The respondent answers by pressing one of
them; each icon has an associated voice recording, in the respondent
language, identifying its function. Speech recognition coverage, in
general, is limited to the specific words and phrases defined in the
application content files. In this particular case, it is advantageous
to limit it further by exploiting the tight constraints inherent in
the questionnaire task, so that at any given moment only the subset of
the coverage relevant to the current question is made available.

As far as rule formalisms are concerned, the questionnaire task only
requires a small extension of the basic translation framework, in order
to add the extra information associated with the questionnaire structure.
The details are straightforward and are described in \cite{ArmandoEA2014}.

The next section uses an initial prototype of a voice questionnaire
app (``AidSLT'') to perform a simple evaluation of speech recognition
performance. The questionnaire used for the evaluation contained 18
fields, which together supported 75 possible translated questions,
i.e.\ an average of about 4 translated questions per field. The recognition grammar
permitted a total of 11\,604 possible source language questions, i.e.\ an average
of about 155 source language questions per translated question.

%

\section{Initial evaluation}
\label{Section:PrototypesEvaluations}

The initial AidSLT questionnaire was tested during the period
March--July 2015 by seven humanitarian workers with field experience
and knowledge of household surveys. The main focus of the
evaluation was on the recognition of speech input by English-speaking
interviewers. Subjects were presented with a simulation exercise that
consisted in administering a household survey about malaria preventive
measures to an imaginary French-speaking respondent.  Instructions
were sent by e-mail in the form of a PDF file. The subjects logged in
to the application over the web from a variety of locations using
password-protected accounts. Each subject ran the questionnaire
once; annotations were added in the script so that several questions 
produced a popup which asked the subject to rephrase their initial
question. 

We obtained a total of 137 correctly logged interactions\footnote{One subject
misunderstood the instructions, one had severe audio problems with their
connection, and a few utterances were spoiled by incorrect use of the push-to-talk
interface.}, which were annotated
independently by two judges. Annotators were asked to transcribe the
recorded utterances and answer two questions for each utterance:
a) whether the subject appeared to be reading the heading for the 
questionnaire field or expressing themselves freely, and b) whether
the translation produced adequately expressed the question asked
in the context of the questionnaire task. Agreement between the two
judges was very good, with a Cohen's kappa of 0.786 and an Intraclass
Correlation Coefficient of 0.922. 

The bottom-line result was that between 77 and 79 of the sentences
were freely expressed (56--58\%) and only 10 produced incorrect
translations (7\%), despite a Word Error Rate of about 29\%. All
the incorrect translations were of course due to incorrect recognition.
We find this result encouraging; the architecture appears to be
robust to bad recognition and exploits the constrained
nature of the task well. 

\section{Conclusions and further directions}

We have described a platform that supports rapid development of a
variety of limited-domain speech translation applications.
Applications can be deployed on the web and run on both desktop and
mobile devices. The minimal formalism is designed to be used by domain
experts who in general will not be computer scientists.

Although the translation platform is still at an early stage of
development, experiences so far are positive; comparisons with our
spoken CALL platform \cite{RaynerEASlate2015Platform}, which uses the same
recognition architecture and has already been tested successfully on a
large scale, leave us optimistic that we will achieve similar results
here. 

\bibliographystyle{splncs03}
\bibliography{FETLT2015}

\end{document}